\documentclass[a4paper, 11pt, fleqn]{article}
\usepackage{a4wide}
\usepackage{amsfonts}
\usepackage{amssymb}
\usepackage{graphicx}
\usepackage{epsfig}
\usepackage{color}

\begin{document}

\title{Anisotropic drop morphologies on corrugated surfaces}

\author {H. Kusumaatmaja\footnotemark[1] , R. J. Vrancken\footnotemark[2] , C.W.M. Bastiaansen\footnotemark[2] , and J. M. Yeomans\footnotemark[1] 
\\
\\ \footnotemark[1] The Rudolf Peierls Centre for Theoretical Physics, Oxford University, 
\\ 1 Keble Road, Oxford OX1 3NP, U.K. and
\\ \footnotemark[2] Chemical Engineering and Chemistry Department, 
\\ Eindhoven University of Technology, Eindhoven 5600 MB, The Netherlands}
\date{\today}


\maketitle

\begin{abstract}

The spreading of liquid drops on surfaces corrugated with micron-scale parallel grooves is studied both 
experimentally and numerically. Because of the surface patterning, the typical final drop shape is no longer 
spherical. The elongation direction can be either parallel or perpendicular to the direction of the grooves, depending on the 
initial drop conditions. We interpret this result as a consequence of both the anisotropy of the contact line 
movement over the surface and the difference in the motion of the advancing and receding contact lines. Parallel to 
the grooves, we find little hysteresis due to the surface patterning and that the average contact angle 
approximately conforms to Wenzel's law as long as the drop radius is much larger than the typical length scale of the 
grooves. Perpendicular to the grooves, the contact line can be pinned at the edges of the ridges leading to large contact 
angle hysteresis. 

\end{abstract}


\section{Introduction}

In recent years wetting and spreading phenomena have received continued attention from the scientific community due 
to their broad application in, for example, microfluidic devices, surface coating and biomimetics. Surface roughness 
can be exploited to significantly alter the behaviour of fluids moving over a surface. Examples include plants where 
micron-scale bumps on the leaves lead to superhydrophobic behaviour \cite{Neinhuis}, desert beetles who use 
hydrophilic patches on their back to collect dew \cite{Parker}, and
butterfly wings which are patterned anisotropically to promote directional run-off \cite{Zheng}. 

It is now possible to reproduce heterogeneous surface patterning in a very controlled manner on micron length scales. 
Regular arrays of chemical patches \cite{Fuchs,Morita,Dupuis1} and posts \cite{Oner,Bico,Lipowsky,Yoshimitsu} of 
different shapes and sizes are regularly fabricated and several authors \cite{Zhu, Gao, Ming} have even 
shown the possibilities of manufacturing multi-scale surface patterns. Recently such patterning has been used to 
control the movement of drops \cite{Kusumaatmaja2,Huck} and to attempt to enhance flow in microchannels 
\cite{Ou,Bocquet}.

The motion of drops on patterned surfaces is complex because of pinning and hysteresis. In particular surfaces with 
anisotropic patterning on a scale comparable to the drop size can result in elongated drop shapes and in a different 
motion parallel and perpendicular to the grooves. Gleiche {\it{et al.}} \cite{Fuchs} showed that there is an 
anisotropy in the average value of contact angle and contact angle hysteresis on chemically nanostructured surfaces. 
Brandon {\it{et al.}} \cite{Marmur} investigated the effect of drop size on chemically striped surfaces and found 
that the drop anisotropy and contact angle hysteresis depended on the drop volume. Elongated drop shapes were also 
obtained by Chen {\it{et al.}} \cite{Patankar} and Chung {\it{et al.}} \cite{Chung} for hydrophobic and hydrophilic 
grooved surfaces respectively. Narhe and Beysens \cite{Narhe} studied the growth dynamics of water drops
condensing on grooved surfaces and showed that similar elongated drop shapes can be found during growth when the
surface is hydrophilic, but is absent when the surface is superhydrophobic. Pakkanen and Hirvi \cite{Pakkanen} further 
showed that the anisotropy still persists when surface patterning is nanoscopic. Morita {\it{et al.}} \cite{Morita} and 
Yoshimitsu {\it{et al.}} \cite{Yoshimitsu} studied the dynamics of drops sliding on chemically striped and hydrophobic 
grooved surfaces and found that the sliding angles are considerably larger for drops moving perpendicular to the stripes.

In this paper we present a number of experiments and lattice Boltzmann simulations of drops spreading or dewetting
on a hydrophilic surface patterned with parallel grooves. Our results highlight the importance of hysteresis
and energetic barriers due to the surface patterning: because of the asymmetric behaviour of the advancing 
and receding contact line, the final drop shape can be elongated either along or perpendicular to the grooves. 
When the contact line is advancing, the drop is found to be elongated parallel to the grooves. On the other hand, when the 
contact line is receding, the drop is elongated perpendicular to the grooves. We present both equilibrium and quasi-static 
drop experiments and simulations in section 4. In section 5, an investigation of the influence of three dimensionless surface 
parameters is presented: the roughness factor, the aspect ratio of the barriers and the dimensions of the drop versus the 
dimensions of the barriers. As expected, the contact angle, particularly that of the interface lying perpendicular to the 
grooves, strongly depends on both the parameters of the surface and the history of the drop motion. The complementary 
nature of experiments and simulations is also demonstrated. Sections 2 and 3 summarise the experimental and numerical 
approaches respectively. 


\section{Experiments}

Two sets of experiments were performed with anisotropically grooved surfaces.

    \begin{figure}
    \centering 
    \includegraphics[width=0.45\textwidth]{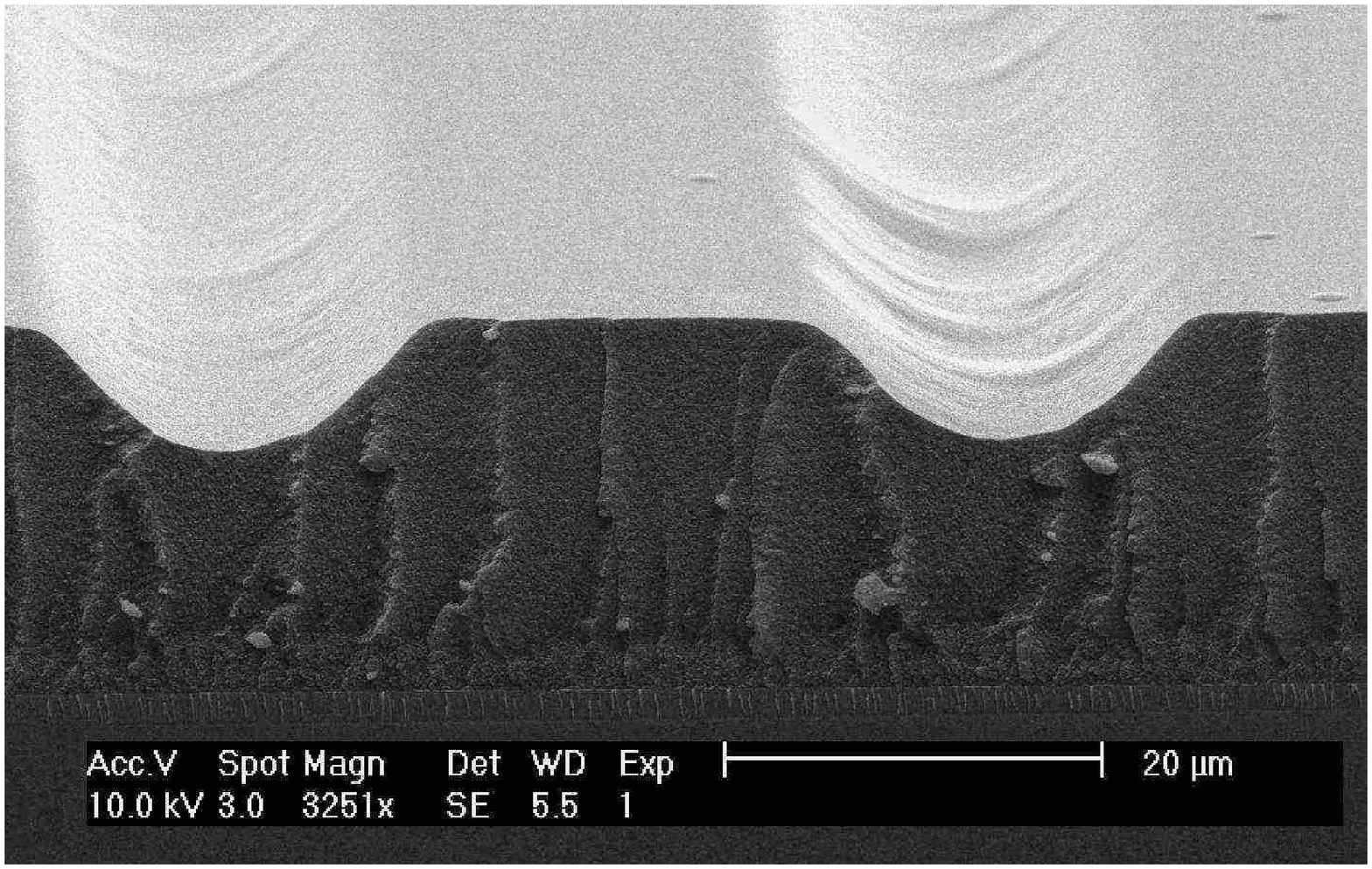}
    \includegraphics[width=0.475\textwidth]{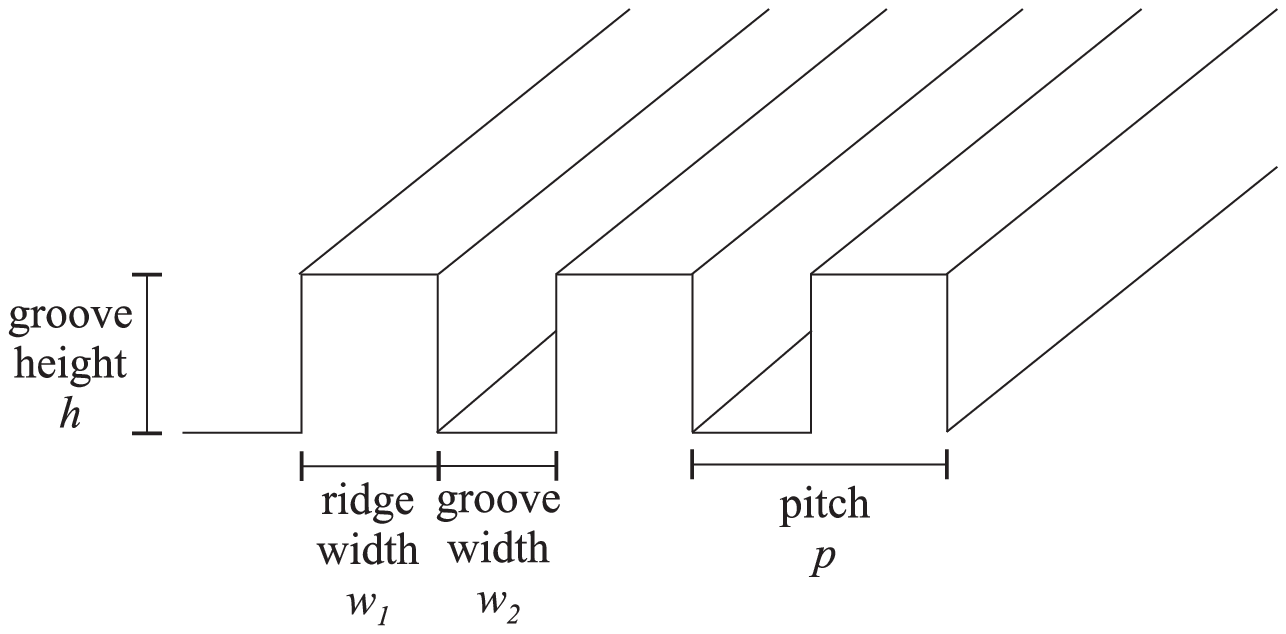}\\
    \caption{\footnotesize Periodically corrugated surfaces used in the (a) experiments and (b) simulations.}
    \label{PESEMKO-SurfSchematic}
    \end{figure}

In the first set of experiments, micron-sized droplets were placed on sub micron-scale corrugated polyimide surfaces. 
The polyimide layers were created on glass by spincoating a commercially available mixture (JSR AL-3046) of polyimide 
from solvent. After spincoating, the samples were heated for 5 minutes at $80^{\mathrm{o}}{\mathrm{C}}$ on a hotplate 
to evaporate the solvent, and afterwards thermally cured in a vacuum oven at $180^{\mathrm{o}}{\mathrm{C}}$ for 1.5 
hours. Afterwards the samples were rubbed unidirectionally with a velvet cloth with a force $\sim 1 \, \mathrm{kg/cm^2}$ 
over a distance of 30 cm with a velocity of 10 cm/s. This created grooves that were determined by AFM to be between 
15 to 20 nm deep and 50 to 200 nm wide. Non-rubbed surfaces were prepared identically, only the rubbing step was omitted.

Surface energy measurements were made on the prepared polyimide surfaces. By measuring the contact angle of water, 
ethylene glycol and diiodomethane drops on the surface, the surface energy was calculated to be 43.6 mN/m, employing the 
OWRK method.

Drops consisting of a monomer of 99.0 w\% ethoxylated bisphenol-A dimethacrylate (Sartomer Europe BV), 
having a surface tension of 41 mN/m, together with 1.0 w\% photo-initiator (Irgacure 184, Ciba Specialty Chemicals), 
were placed on the surface by two methods. In the first method, the mixture was 
sprayed manually onto the polyimide surface. The resulting drops were placed randomly on the surface and were of 
varying size. In the second method, micro-transfer printing \cite{Sanchez}, a PDMS elastomeric stamp with an array 
of trapezoidal posts with face areas of 100 x 100 $\mu$m was pressed into a thin spincoated layer of monomer mixture and afterwards pressed 
against the polyimide layer. Part of the mixture was transferred to the polyimide surface, after which the stamp was 
detached from the surface leaving an array of drops with equal size. The drops were polymerized 
within one minute after placement. This fixed the geometry of the drop and enabled analysis afterwards, while still 
allowing the drop sufficient time to equilibrate. The drop $x-y-z$ profiles were determined by optical interferometry 
(Fogale Zoomsurf 3D), and the parallel and perpendicular cross-sectional height profiles were extracted and 
ellipse-fitted to determine the contact angles and elongation. A minimum of five drops were examined for each surface. 

In the second set of experiments, smooth corrugated polymer surfaces were created by means of a four-step process 
called photo-embossing \cite{Hermans}. First a so-called photo-polymer layer was spin-coated on glass forming layers 
approximately 20 $\mu$m thick. The mixture consisted of 41.9 w\% monomer dipentaerythritol penta/hexa-acrylate 
(Sigma Aldrich), 41.9 w\% polymer polybenzylmethacrylate (Mw 70 kg/mol; Scientific Polymer Products), 4.20 w\% 
photo-initiator Irgacure 819 (Ciba Specialty Chemicals) and 12.0 w\% inhibitor tert-butyl hydroquinone (Aldrich). In the 
second step, the layer was illuminated with UV light through a suitable lithographic mask (described below), partially 
polymerizing the exposed parts of the layer. In the third step, the sample was heated and part of the monomers in the 
non-illuminated regions diffused towards the illuminated regions, creating the corrugated surface. In the last step the 
surface was flood-exposed with UV light to polymerize the full layer. Similarly to the polyimide surface, the 
photo-embossed surface was found to have a surface energy of 44.7 mN/m.

The lithographic masks employed had sizes of 1 x 1 cm and consisted of parallel stripes, where each stripe was as wide 
as the distance between the stripes, i.e. 50 \% of the incident light is transmitted. The pitch, i.e. the repeat length of the 
mask (line to line) was varied from 10 $\mu$m to 100 $\mu$m. The depth of the grooves could be varied between 100 nm 
and 5 $\mu$m by varying the illumination time, with the exception that the maximum depth could only be reached with 
masks with pitches above 40 $\mu$m because sufficient material needs to available to be be displaced by the 
photo-embossing process. The depth of the grooves was again determined with optical interferometry.

Water drops of 4 $\mu l$ were placed on the photo-embossed polymer surfaces by an automated dispensing needle 
(Optical Contact Angle setup, Dataphysics) at a constant height above the surface to minimize any variations in drop 
placement. The water used for the measurements was purified (Millipore Super-Q) and CO$_{2}$ was removed (Elix-10
UV). The drop properties were measured on the surface with the contact angle measurement setup . Both $\theta_{\parallel}$ 
and $\theta_{\perp}$ were determined, as well as the drop elongation $e$. The contact angles and base radii were determined by means 
of digital drop shape fitting employing an ellipse fitting algorithm. 

The two sets of experiments are complementary, in that they employ different surfaces and probe liquids. The rubbed surfaces have nano-meter scale corrugations, whereas the photo-embossed surfaces have micron-scaled corrugations. Micro-transfer printing is suitable for creating well-defined dewetting drops, while photo-embossed surfaces have the great benefit of being smooth as well as controllable with regards to the size of the corrugations. In the following sections the results from both sets of experiments are presented next to each other highlighting different aspects of anisotropic hysteresis.


\section{Lattice Boltzmann simulations}

We use a mesoscale simulation approach where the equilibrium properties of the drop are modelled by a continuum free 
energy
    \begin{equation} 
    \Psi = \int_V (\psi_b(n)+\frac{\kappa}{2} (\partial_{\alpha}n)^2) dV
    + \int_S \psi_s(n_s) dS . 
    \label{eq3}
    \end{equation}   
$\psi_b(n)$ is a bulk free energy term which we take to be \cite{Briant1}
    \begin{equation}
    \psi_b (n) = p_c (\nu_n+1)^2 (\nu_n^2-2\nu_n+3-2\beta\tau_w) \, ,
    \end{equation}
where $\nu_n = {(n-n_c)}/{n_c}$, $\tau_w = {(T_c-T)}/{T_c}$ and $n$, $n_c$, $T$, $T_c$ and $p_c$ are the local 
density, critical density, local temperature, critical temperature and critical pressure of the fluid respectively. 
This choice of free energy leads to two coexisting bulk phases (liquid and gas) of density 
$n_c(1\pm\sqrt{\beta\tau_w})$. The second term in Eq.\ (\ref{eq3}) models the free energy associated with any 
interfaces in the system. $\kappa$ is related to the surface tension via $\gamma = {(4\sqrt{2\kappa p_c} 
(\beta\tau_w)^{3/2} n_c)}/3$ \cite{Briant1}. The last term in Eq.\ (\ref{eq3}) describes the interactions between the 
fluid and the solid surface. Following Cahn \cite{Cahn} the surface energy density is taken to be $\psi_s (n) = -\phi 
\, n_s$, where $n_s$ is the value of the fluid density at the surface. The strength of interaction, and hence the 
local equilibrium contact angle, $\theta_e$, is parameterized by the variable $\phi$. Here the simulation parameters 
are chosen to give $n_{\mathrm{liquid}} = 4.11$, $n_{\mathrm{gas}} = 2.89$, $\gamma = 5.14 \, \times \, 10^{-4}$, and 
$\theta_e = 70^{\mathrm{o}}$. The typical drop and pattern sizes are of order 100 and 10 lattice spacings respectively.  
Simulation and physical parameters are related by choosing a length scale $l_0$, a time scale $t_0$, and a mass 
scale $m_0$. A simulation parameter with dimensions $[l]^{n1}[t]^{n2}[m]^{n3}$ is multiplied by 
$[l_0]^{n1}[t_0]^{n2}[m_0]^{n3}$ to give the physical value \cite{Dupuis3}. More specific details on the model used here, 
including the way we have implemented complicated surface geometries, can be found in \cite{Dupuis2}. 

The dynamics of the drop is described by the continuity  (\ref{eq1}) and the Navier-Stokes equations (\ref{eq2})
    \begin{eqnarray}
    &\partial_{t}n+\partial_{\alpha}(nu_{\alpha})=0 \, , 
    \label{eq1}\\
    &\partial_{t}(nu_{\alpha})+\partial_{\beta}(nu_{\alpha}u_{\beta}) = 
    - \partial_{\beta}P_{\alpha\beta}+ \nu \partial_{\beta}[n(\partial_{\beta}u_{\alpha} + \partial_{\alpha}u_{\beta} 
+ \delta_{\alpha\beta} \partial_{\gamma} u_{\gamma}) ] \, , 
    \label{eq2}
    \end{eqnarray}
where $\mathbf{u}$, $\mathbf{P}$, and $\nu$ are the local velocity, pressure tensor, and kinematic viscosity 
respectively. The thermodynamic properties of the drop are input via the pressure tensor $\mathbf{P}$ which is 
calculated from the free energy \cite{Briant1,Dupuis2,Dupuis3} via
    \begin{eqnarray} 
    &P_{\alpha\beta} = (p_b(n)-\frac{\kappa}{2} (\partial_{\alpha}n)^2 -\kappa n 
\partial_{\gamma\gamma}n)\delta_{\alpha\beta} 
    + \kappa (\partial_{\alpha}n)(\partial_{\beta}n) \, , \\ 
    &p_b (n)= p_c (\nu_n+1)^2 (3\nu_n^2-2\nu_n+1-2\beta\tau_w) . 
    \end{eqnarray}
Eqs. (\ref{eq1}) and (\ref{eq2}) are solved using a lattice Boltzmann algorithm 
\cite{Briant1,Dupuis2,Dupuis3,Swift1,Succi}. 

This model (or very similar approaches e.g. \cite{Kwok,Succi2,Timonen}) has been shown to be a useful tool to study 
several aspects of drop dynamics on patterned surfaces, such as the problem of contact angle hysteresis 
\cite{Kusumaatmaja}, and drops spreading on superhydrophobic \cite{Dupuis2} and chemically patterned surfaces 
\cite{Dupuis3}. 

To render the simulation feasible, the typical surface pattern used in the simulations is slightly different to that 
produced experimentally. It is therefore important to note here that our aim in this paper is to obtain a consistent 
qualitative understanding of the problem rather than to attempt to exactly match experiments and simulations. The 
shape of the grooves is taken to be rectangular, as shown in Fig. \ref{PESEMKO-SurfSchematic}(b). We define the 
aspect ratio as the ratio of the groove height to the pitch. No--slip boundary conditions are imposed on the velocity 
field on the surfaces adjacent to and opposite to the drop and periodic boundary conditions are used in the other two 
directions. 

\section{Anisotropic drop morphology on a corrugated surface}

In this section, the anisotropy in the final shape of a drop spreading on a corrugated surface is investigated. First, we performed experiments where liquid drops are sprayed or micro-transfer printed onto the rubbed polyimide surface. 
We find that even though both preparation methods result in an elongated drop shape, the direction of elongation relative to the surface 
grooves is different. The drop is elongated parallel to the grooves if sprayed, while micro-transfer printed drops are elongated perpendicular to the grooves. 

Results consistent with the first set of experiments above were found when water is quasi-statically added to a spreading drop prepared on
a photo-embossed surface. Here different qualitative behaviours are observed parallel and perpendicular to the grooves, further highlighting
the asymmetry of contact line motion in these two directions. The experiments are then complemented by performing 
lattice Boltzmann simulations to further understand the role of the surface anisotropy in determining the final shape of the drop.

    \begin{figure}
    \centering 
    \includegraphics[width=0.95\textwidth]{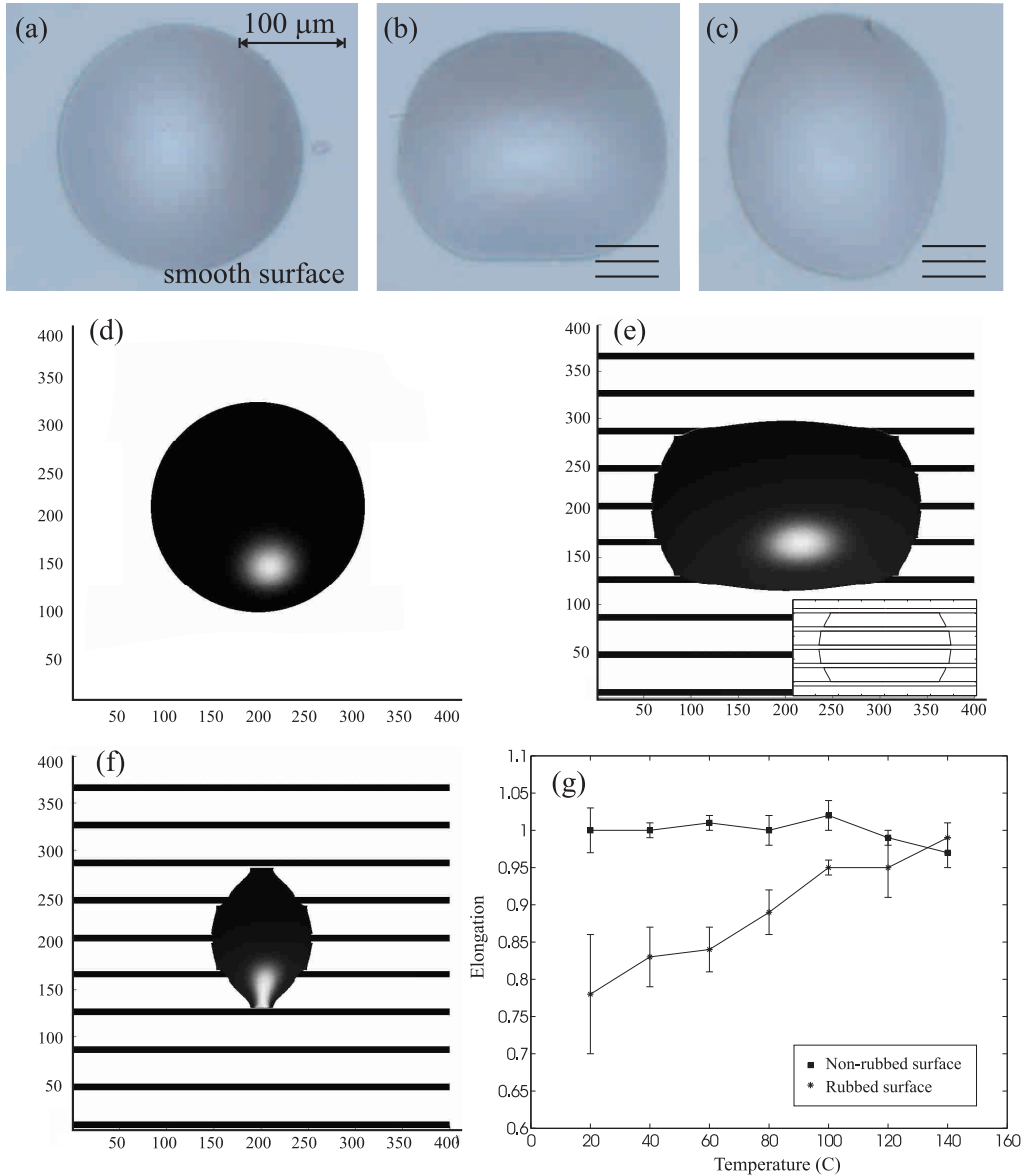}\\
    \caption{\footnotesize Typical drop shape observed in experiments when (a) there is no surface patterning, 
    (b) the drop is sprayed onto the surface and spreads, and (c) the drop is stamped onto the surface and dewets. 
    Although not visible in this image, the surfaces of (b) and (c) were patterned with sub micron-scale ridges. 
    Here the ridges are oriented in the horizontal direction. (d), (e) and (f) Lattice Boltzmann simulation results 
    when (d) the drop spreads on a smooth surface, (e) and (f) the drop volume is quasi-statically (e) increased 
    and (f) decreased. The experimental drop elongation parallel (b, e) and perpendicular (c, f) to the grooves is 
    clearly reproduced in the simulations. (g) Experimental measurements of the average elongation as function 
    of printing temperature for the printed acrylate drops on rubbed and non-rubbed polyimide. The drops on 
    non-rubbed surfaces have approximately spherical contact lines at all temperatures, whereas with decreasing 
    temperatures the drops on rubbed surfaces become more and more elongated.}
    \label{PrintedDrops}
    \end{figure}

\subsection{Experiments}

In Fig. \ref{PrintedDrops}(b) we show a typical final shape of a liquid drop that is sprayed onto the surface, causing 
the the drop contact line to advance to wet the corrugated surface. Since the surface patterning is not isotropic, the 
advancing contact line behaves differently parallel and perpendicular to the grooves. Based on Johnson and Dettre's 
work \cite{Johnson} no hysteresis due to the surface patterning is expected in the parallel direction as 
there are no energetic barriers present which could tend to pin the contact line. On the other hand, perpendicular to 
the grooves, surface undulations are known to pin the contact line \cite{Huh2,Gibbs,Bailey}. Contact line pinning leads to 
strong contact angle hysteresis and this causes the advancing contact angle, $\theta_A$, to differ from that in the 
parallel direction. 

\begin{figure}
    \centering 
    \includegraphics[width=0.75\textwidth]{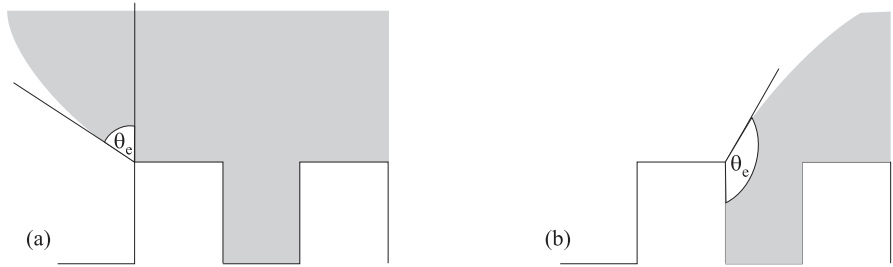}\\
    \caption{\footnotesize Graphical illustration of Gibbs criteria: (a) advancing and (b) receding contact line.}
    \label{GibbsCriteria}
    \end{figure}

The pinning is illustrated in Fig. \ref{GibbsCriteria}(a). For the drop contact line to advance, it has to wet the  
side of the grooves, which, according to the Gibb's criterion \cite{Gibbs,Bailey}, occurs when $\theta_A = \theta_e + 
90^{\mathrm{o}}$ for rectangular ridges. More generally $\theta_A = \theta_e + \alpha$ in two dimensions where 
$\alpha$ is the maximum inclination of the surface. In three dimensions, the value of the advancing (and receding - 
see below) angle is not extreme as that predicted by the Gibbs' criteria, due to the energy costs associated with the 
surface deformation from the spherical cap shape. The advancing angle is, nonetheless, generally larger than the 
advancing angle parallel to the grooves. As a consequence, it is easier for the drop to spread in the parallel 
direction and hence the drop shape is elongated parallel to the grooves as shown in Fig. \ref{PrintedDrops}(b). 
This is the final drop shape typically found in corrugated surface experiments \cite{Patankar,Pakkanen,Chung}. 
The shape of a drop on a smooth surface is also shown in Fig. \ref{PrintedDrops}(a) for comparison.

When the drop is (micro-transfer) printed onto the surface, on the other hand, the drop contact line retreats to dewet the surface, as the liquid is initially spread out much further than its equilibrium shape. 
This occurs when the drop contact angle is smaller than or equal to the receding contact angle. If no contact angle 
hysteresis is present, the final drop shape will be independent of the initial conditions. However, as discussed 
above, hysteresis is an important effect in the perpendicular direction and as a result, the receding angles are 
again different parallel and perpendicular to the grooves. Contact line pinning for the receding motion is 
illustrated in Fig. \ref{GibbsCriteria}(b). For the contact line to recede, the drop has to dewet the side of 
the posts, which happens when $\theta_R = \theta_e - 90^{\mathrm{o}}$ for rectangular ridges. For other geometries, 
$\theta_R = \theta_e - \alpha$ in two dimensions where $\alpha$ is the maximum inclination of the surface. Since the 
receding angle in the perpendicular direction is smaller than the receding angle parallel to the grooves, it is 
easier for the drop to dewet in the parallel direction and the final drop shape is typically elongated perpendicular 
to the grooves, as shown in Fig. \ref{PrintedDrops} (c).

    \begin{figure}
    \centering
    \includegraphics[width=1.0\textwidth]{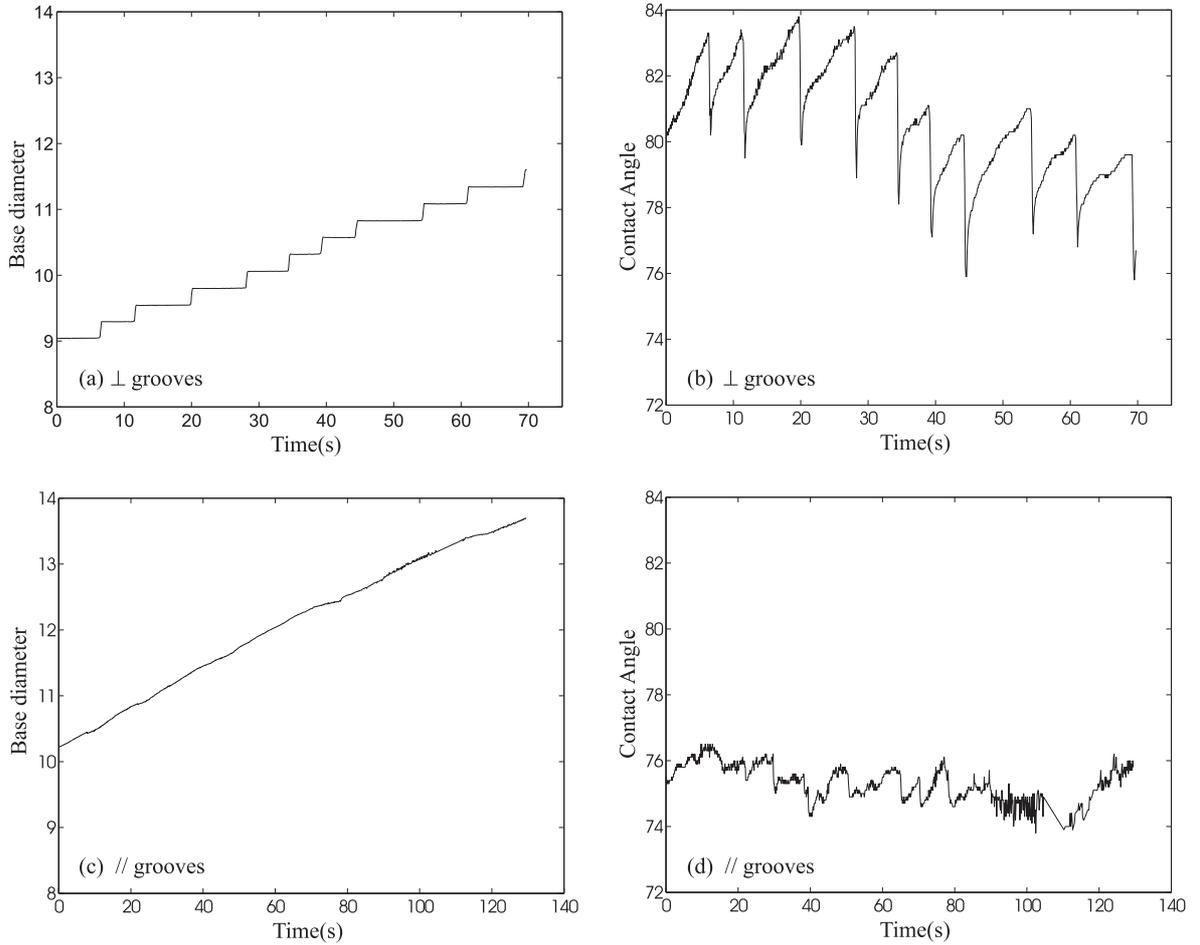}\\
    \caption{\footnotesize Advancing contact angle measurement for a drop of water spreading on a photo-embossed surface. (a) and (c) are measurements of the base diameter (in arbitrary units because the setup was not length-calibrated), while (b) and (d) are the contact angles as a function of time. (a) and (b) are the diameter and contact angle 
perpendicular to the grooves, (c) and (d) show the same variables parallel to the grooves. The initial drop volume was 
 4 $\mu$l. Water was added with 0.06 $\mu$l per second effecting a quasi-static, linear increase in volume. The 
surface had a pitch 80 $\mu$m and groove depth of 4.3 $\mu$m, i.e. an aspect ratio of 0.10. The equilibrium contact 
angle of the surface was $\theta_e \simeq 70^{\mathrm{o}}$. The characteristic effect of crossing surface barriers in the perpendicular direction is clearly visible in (a) and (b), while (c) and (d) demonstrate that in the parallel direction the motion is much smoother.}.
    \label{AdvRecPE}
    \end{figure} 

The effect of the printing temperature on the final shape of the printed drops showed an interesting result, as shown in 
Fig. \ref{PrintedDrops}(g). With increasing temperature, the elongation perpendicular to the barriers decreases. Clearly, 
the pinning ability of the barriers is decreased at higher temperatures and the drop morphology is less affected by hysteresis. 
A possible explanation for this effect is that the viscosity at higher temperature decreases so that the drop in general and the contact line in particular move faster 
and dissipate less energy. Even without determining the exact barrier crossing mechanism, in general the interface can more 
easily overcome pinning on the ridges by having more energy available for deforming its surface locally, enabling it to cross 
more barriers. This is further supported by our simulation results, where we found similar dependence of drop elongation on 
the drop viscosity and impact velocity. The typical drop elongation decreased as we reduced the drop viscosity or increased 
the drop initial kinetic energy.

\begin{figure}
    \centering
    \includegraphics[width=0.75\textwidth]{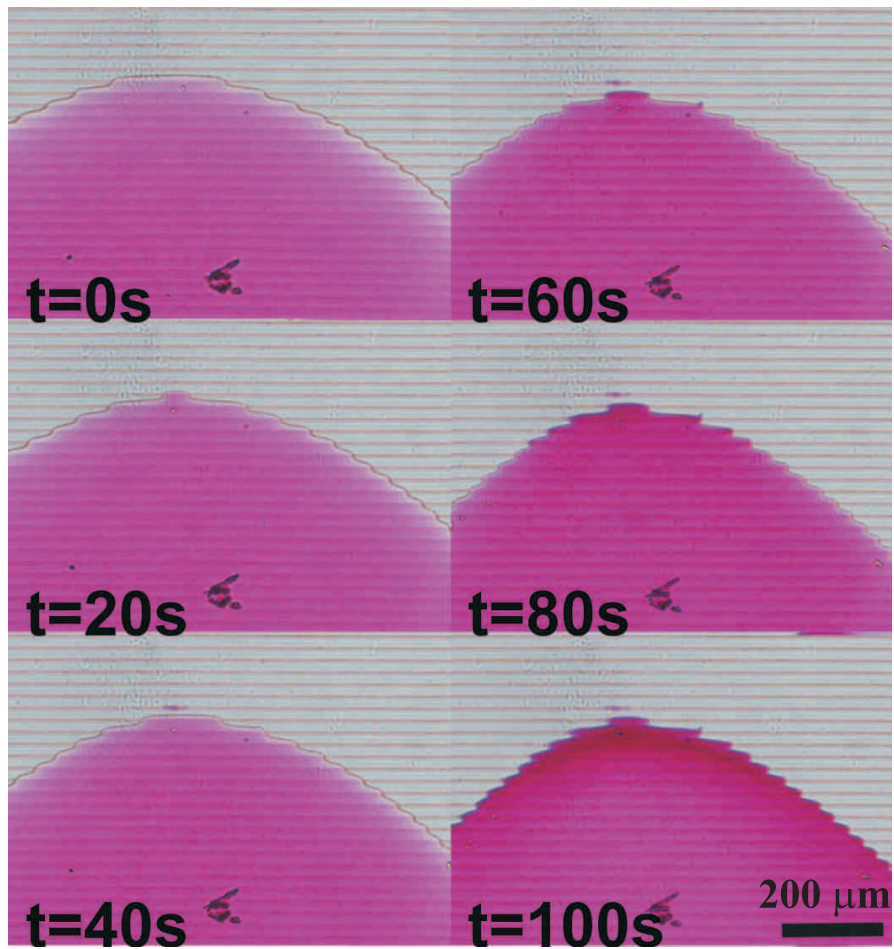}\\
    \caption{\footnotesize (color online) Microscope observation of the shape of the receding contact line as a water drop evaporates 
    over time. The drop retreats parallel to the grooves until it becomes sufficiently distorted that it is favourable for it to retract, dewetting 
    the outermost grooves. 1\% Rhodamine B is added to the drop for visibility.}
    \label{receding}
    \end{figure}

The mechanism by which the interface crosses the ridges can be envisaged as the nucleation 
mechanism formulated by DeGennes \cite{deGennes}: in this process an advancing drop will first form a nucleus in the next
grooves, after which the interface spreads along the grooves until a new and stable, but less elongated morphology is reached. 
For drops retracting over the barrier, the inverse process occurs. The drop retreats parallel to the grooves until it becomes
sufficiently distorted that it is favourable for it to retract, dewetting the outermost grooves. We were able to observe this retracting 
mechanism with evaporating water drops which were placed on a photo-embossed surface. 
This is shown in Fig. \ref{receding}.

Our interpretation of the role of contact line pinning in determining the drop shape is backed up by the experimental
evidence obtained when considering a quasi-statically growing drop on a photo-embossed surface, shown in Figs. 
\ref{AdvRecPE}. The drop contact angles and base radii both parallel and perpendicular to the grooves are measured 
as a function of drop volume. Three distinct features are visible: (i) the saw-tooth shaped variation with time of the 
perpendicular contact angle, (ii) the step-shaped variation of the perpendicular base radius and (iii) the absence of these 
distinct features in the parallel direction where the base radius increases continuously and the contact angle is 
roughly constant over the whole range of the experiment. Features (i) and (ii) are clear indications of contact line 
pinning in the perpendicular direction, while feature (iii) shows that pinning of the contact line due to the 
surface patterning does not occur in the parallel direction, leading to a significantly lower advancing contact angle. 
These features were reproduced for all aspect ratios and surfaces. For receding contact line, no depinning was observed 
in the parallel direction, while depinning in the perpendicular direction occurred irregularly, with interfaces sometimes 
crossing multiple ridges at once. For this reason, reproducibly quantifying the depinning as function of the surface 
parameters was unsuccessful. We found that, on the unpatterned surfaces, receding contact angle measurements 
give values of about $15^{\mathrm{o}}$ to $25^{\mathrm{o}}$ with occasional outliers of up to $50^{\mathrm{o}}$. 
On the patterned surfaces, receding contact angle measurements in the parallel direction give similar values to those 
for the unpatterned samples, while the receding angles in the perpendicular direction are on average about $5^{\mathrm{o}}$ 
lower than those in the parallel direction. The high experimental scatter is attributed to the common problem of 
measuring very low receding contact angles, namely that incidental pinning and depinning are observed even on 
(apparently) smooth surfaces. The base radii were also greater in the perpendicular direction, further proving 
that hysteresis is increased by the barriers.
    	
\subsection{Simulations}
	
Similar hysteresis in the motion of the drop perpendicular to the grooves is seen in lattice Boltzmann 
simulations. A liquid drop was allowed to equilibrate on the corrugated surface. Then, using this as an 
initial condition, we ran two sets of simulations where the drop volume was quasi-statically increased in the first 
set and decreased in the second set. To do this, we varied the liquid density by $\pm 0.1\%$ from its equilibrium 
bulk density every 10000 time steps. This changed the drop volume as the system relaxed back to its coexisting equilibrium densities. 

    \begin{figure}
    \centering 
    \includegraphics[width=0.45\textwidth]{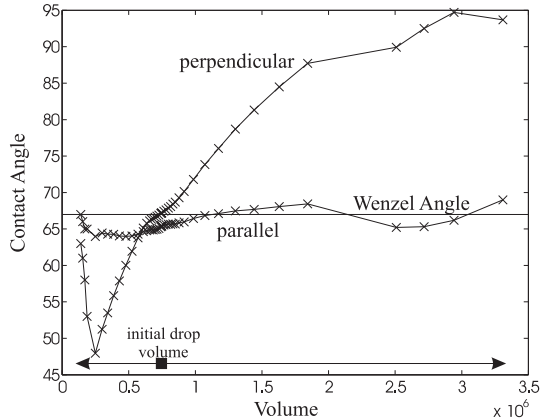}\\
    \caption{\footnotesize Simulations of drop contact angles on a grooved surface as a function of the drop volume 
    for groove height $h = 3$, ridge width $w_1 = 10$ and pitch $p = 40$. The drop was initialized at $V\sim0.75 \times 10^6$ 
    and equilibrated. Then the drop volume was either slowly decreased or increased.}
    \label{SimResults2}
    \end{figure}

When the drop volume is increased, the drop contact line advances. Therefore, we expect the first set of simulations 
to mimic a drop sprayed onto the surface and the final drop shape to be elongated parallel to the grooves. When the 
drop volume is decreased, to the contrary, the drop contact line recedes. We therefore expect the second set to mimic 
a drop stamped onto the surface and the drop shape to be elongated perpendicular to the grooves. The simulation 
results shown in Fig. \ref{PrintedDrops} (d) and (e) clearly show that the drop shape is indeed elongated parallel 
and perpendicular to the grooves when the contact line is (d) advancing and (e) receding. The drop contact angles 
were also tracked during the simulation and the results are shown in Fig. \ref{SimResults2}. The main features 
indicating contact line pinning in the perpendicular direction are again reproduced: (i) the parallel contact angle 
stays roughly constant and (ii) a large variation in the perpendicular contact angle measurements is observed. 
Note, however, that due to numerical limitations, it was only possible to follow the drop as it moved over one 
groove so that Fig. \ref{SimResults2} shows just one period of the repeating sawtooth pattern obtained in the 
experiments in \ref{AdvRecPE}. In the figure the receding contact line jump in the perpendicular direction
occurs at $V \sim 0.3 \times 10^6$.

    
\section{Drops spreading on corrugated surfaces: understanding the contact angle measurements}

In this section we attempt to provide a consistent picture of the geometry of a drop spreading on corrugated surfaces 
over a wide range of surface parameters. In particular, we shed light on the influence of three dimensionless surface 
parameters: the roughness factor, the aspect ratio of the barriers and the relative ratio of the dimensions of the drop 
to the dimensions of the barriers. Our conclusions will be drawn from both experiments and lattice Boltzmann 
simulations.

We have demonstrated that the shape of a drop spreading on a corrugated surface is no longer spherical. The drop is 
elongated either in the direction parallel or perpendicular to the grooves and the contact angle varies along the 
contact line. In order to simplify the analysis, we focus only on two principal directions: parallel and 
perpendicular to the grooves, since the greatest difference in contact angle is expected between these two. We define 
the apparent contact angle as the contact angle made by fitted ellipses at its intersection with the top of the 
ridges. It is very important to realize, however, that the apparent angle is often not equal to the local contact 
angle, in particular where the contact line is distorted locally by the corrugations. We also define the aspect 
ratio as the ratio of the groove height to the pitch and the drop elongation as the ratio of the maximum base radius 
in the parallel and perpendicular directions.

    \begin{figure}
    \centering 
    \includegraphics[width=1.0\textwidth]{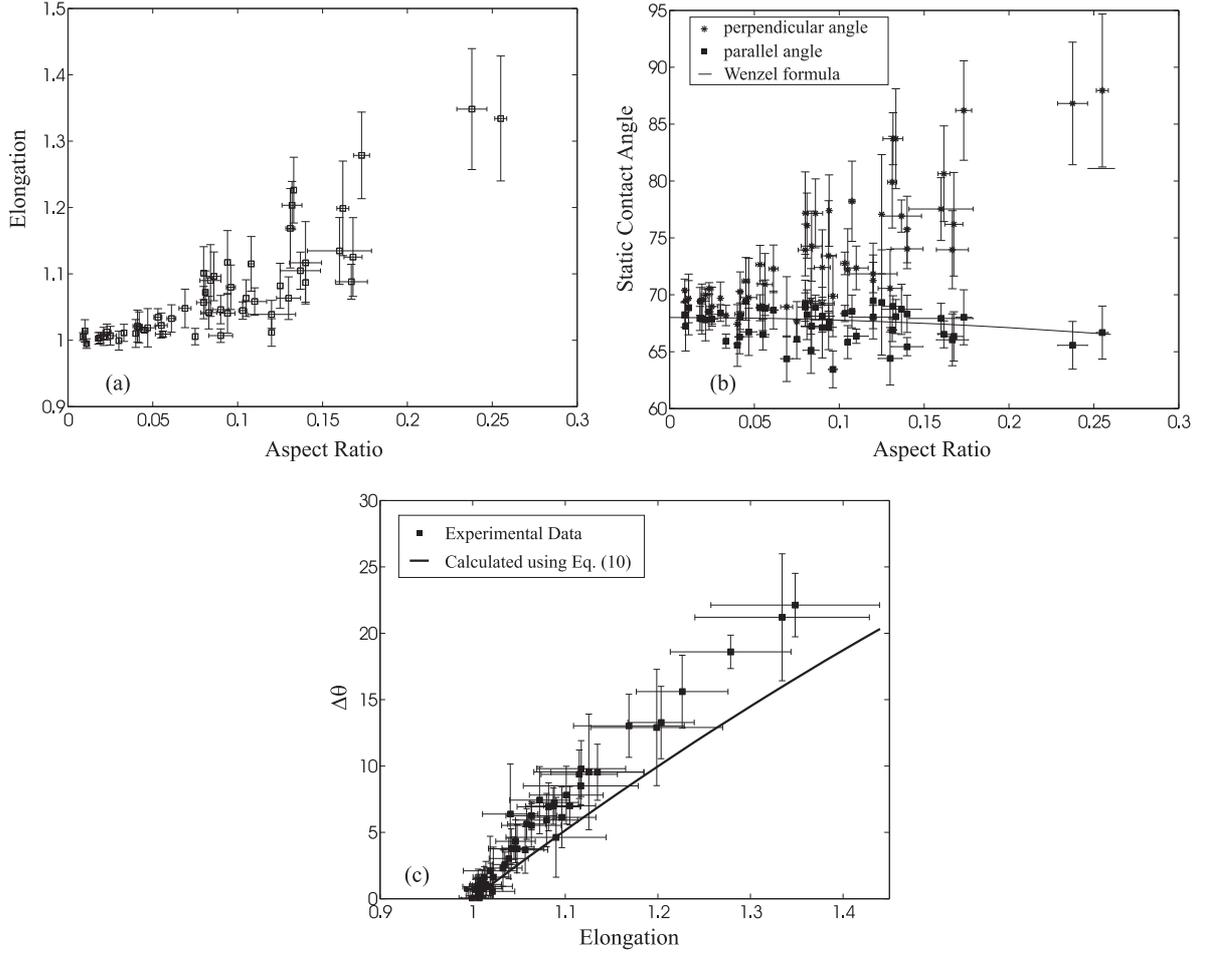}
    \caption{\footnotesize Experimental results for water drops on photo-embossed surfaces. (a) Elongation of the 
droplet  and (b) Static $\theta_{\parallel}$ and $\theta_{\perp}$ as functions of the aspect ratio of the surface. Both the 
elongation and $\theta_{\perp}$ increase approximately monotonically with increasing aspect ratio. The solid line 
corresponds to the Wenzel angle. (c) $\Delta\theta = \theta_{\perp} - \theta_{\parallel}$ as a function of 
elongation. The solid line is the relation between $\Delta\theta$ and the elongation predicted by Eq. \ref{eq7} and 
using $\theta_{\parallel} = \theta_{\mathrm{Wenzel}}$.}
    \label{ElongPerpParAspRatio}
    \end{figure}
	
\subsection{Experiments}

Fig. \ref{ElongPerpParAspRatio}(a) and (b) show our experimental results on how the drop elongation and contact 
angles depend on the aspect ratio. In this set of experiments, we prepared photo-embossed surfaces having different corrugations and 
liquid drops were placed onto the surfaces with a syringe. We first found that there was no obvious relationship 
between the final drop geometry and either the groove depth or the groove pitch. However, plotting the data against 
the aspect ratio of the corrugations, trends could be identified, although the scatter was significant. This is expected given 
the important role of hysteresis and the existence of multiple local energy minima. 

There are four aspects of the data which we wish to comment on: 

(i) On chemically striped surfaces, it has been shown experimentally \cite{Fuchs} and theoretically \cite{Johnson} 
that there is no contact angle hysteresis (at least due to the surface patterning) in the direction parallel to the 
stripes. Furthermore, Gleiche {\it{et al.}} \cite{Fuchs} showed that the parallel contact angle is close to the 
Cassie-Baxter contact angle \cite{Cassie}. Similarly, in this study, we might expect the parallel angle to 
approximately follow the Wenzel formula \cite{Wenzel},
    \begin{equation}
    \cos{\theta_{\parallel}} = r \cos{\theta_{e}}, 
    \label{eq4}
    \end{equation}
where $r$ is the roughness factor. In the experiments
    \begin{equation}
    r \simeq \frac{w_1 + \sqrt{w_2^2 + 4 \, h^2}}{p} \, 
    \label{eqRPE}
    \end{equation}
where $w_1\simeq w_2$. $w_1$, $w_2$, $h$ and $p$ are respectively the ridge width, the groove width, the groove height, and the 
pitch. The equilibrium contact angle $\theta_e \simeq 68^{\mathrm{o}}$. In figure \ref{ElongPerpParAspRatio}(b) 
the Wenzel approximation is indeed shown to correspond well to $\theta_{\parallel}$. The same conclusion was also 
found in \cite{Chung}.

(ii) The perpendicular contact angle, on the other hand, shows a very large scatter. Typically $\theta_{\perp} > 
\theta_{\parallel}$, and $\theta_{\perp}$ tends to increase with the aspect ratio. This may be explained using the 
Gibbs' criteria \cite{Gibbs} we discussed in the previous section. Since the drop spreads outwards on the corrugated surface, it 
is appropriate to consider the advancing contact line motion. Therefore, it is easier for the drop 
to spread parallel to the grooves. This immediately implies that the drop contact angle perpendicular to the grooves is 
larger than that parallel to the grooves. The maximum value of $\theta_{\perp}$  is an estimate of the advancing 
angle in the perpendicular direction. For the two dimensional model illustrated in Fig. \ref{GibbsCriteria}, 
$\theta_A = \theta_e+\alpha$, where $\alpha$ is the maximum slope on the surface. This is clearly an overestimate in 
three dimensions, since it neglects the energy costs associated with the surface deformation from a spherical cap. 
Nevertheless, from the data shown in Fig. \ref{ElongPerpParAspRatio}, we found that the advancing angle increases 
from $70^{\mathrm{o}}$ to $88^{\mathrm{o}}$ with increasing aspect ratio. This is consistent with the fact that, for 
the experimental substrate geometry, (Fig. \ref{PESEMKO-SurfSchematic} (a)), the maximum slope $\alpha$ increases 
with the aspect ratio of the grooves. Similar behaviour is also observed in Fig. 2(e) of a recent work by 
Chung {\it{et al.}}  \cite{Chung}.

(iii) In Fig. \ref{ElongPerpParAspRatio} (c), we plot the difference between the contact angles $\Delta\theta$ as a 
function of the drop elongation. We find that $\Delta \theta$ varies monotonically with drop elongation and the 
experimental scatter is reduced when compared to figure \ref{ElongPerpParAspRatio} (a) and (b). Qualitatively this 
can be explained by assuming that the drop shape is close to elliptical. Assume for simplicity that the drop profiles 
in the two principal directions can be fitted to circles. In this case, the apparent contact angles can be written as 
    \begin{equation}
    \tan{\theta_{\parallel}/2} = h/a$ \qquad \mbox{and} \qquad $\tan{\theta_{\perp}/2} = h/b,
    \label{eq6}
    \end{equation}  
where $a$ and $b$ are the drop base lengths in the two principal directions. As a result, 
    \begin{equation}
    \tan{\theta_{\perp}/2} = e \, \tan{\theta_{\parallel}/2}.
    \label{eq7}
    \end{equation}  
$\theta_{\parallel}$ is approximately constant in the experiments and hence $\theta_{\perp}$ increases monotonically 
with $e$. In Fig. \ref{ElongPerpParAspRatio}(c), Eq. (\ref{eq7}) is plotted assuming $\theta_{\parallel} = 
\theta_{\mathrm{Wenzel}}$, and again it corresponds well to the trend observed in the experiments.

(iv) The drop elongation $e$ increases monotonically with the aspect ratio. This result follows from the fact that 
the energy gain of spreading in the direction parallel to the grooves is higher for higher aspect ratio.

    \begin{figure}
    \centering 
    \includegraphics[width=0.95\textwidth]{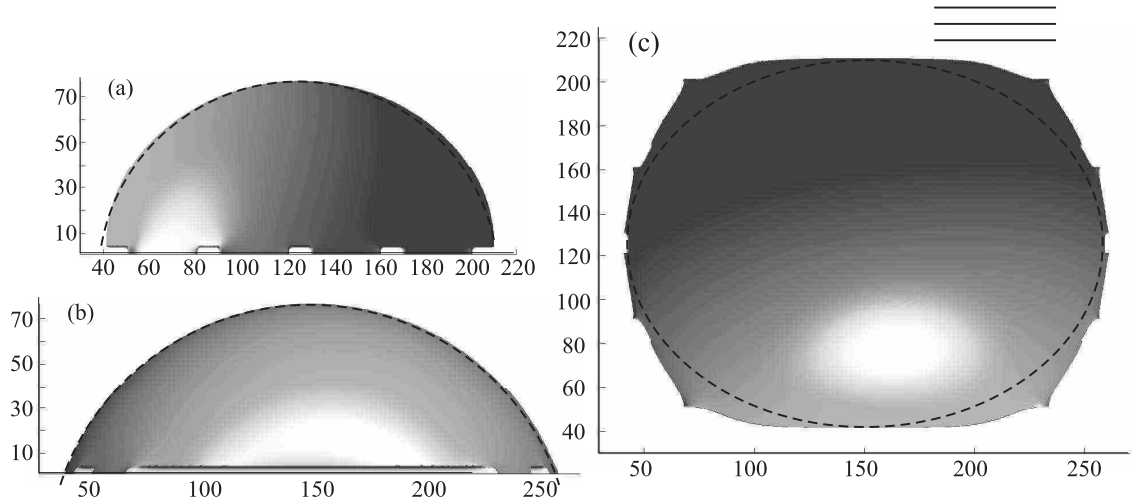}\\
    \caption{\footnotesize A typical drop shape observed in simulations. (a) and (b) the drop cross sections in the 
direction parallel and perpendicular to the grooves. (c) Top view of the drop. The dashed lines are fits to 
(a) and (b) a circle or (c) an ellipse.}
    \label{NumTopSideView}
    \end{figure}

\subsection{Simulations}
	
To gain further insight into these relationships between surface and drop geometry, simulations were performed. The 
drop was initialized as a spherical drop just above the surface and was allowed to spread, without any external 
force, on the corrugated surface. The drop elongation and contact angles were then recorded after 400000 time steps, 
by which time the drop had reached its (meta) stable configuration. A typical final drop shape is shown in Fig. 
\ref{NumTopSideView}. As shown in Fig. \ref{NumTopSideView}(a) and (b) the drop cross sections in the direction 
parallel and perpendicular to the grooves can be fitted well with ellipses. We have also tried to fit the contact 
line around the base of the drop, see e.g. figure \ref{NumTopSideView}(c), but the typical fit we obtain is poor: 
indeed, the top view of the drop clearly shows a corrugated contact line. The corrugation is more noticable with 
increasing aspect ratio. This is due to the fact that it is more advantageous to wet the sides of the grooves the 
higher the aspect ratio. This is similar to changing the wettability ratio between the hydrophilic and hydrophobic stripes
on chemically striped surfaces.

\begin{figure}
    \centering 
    \includegraphics[width=0.9\textwidth]{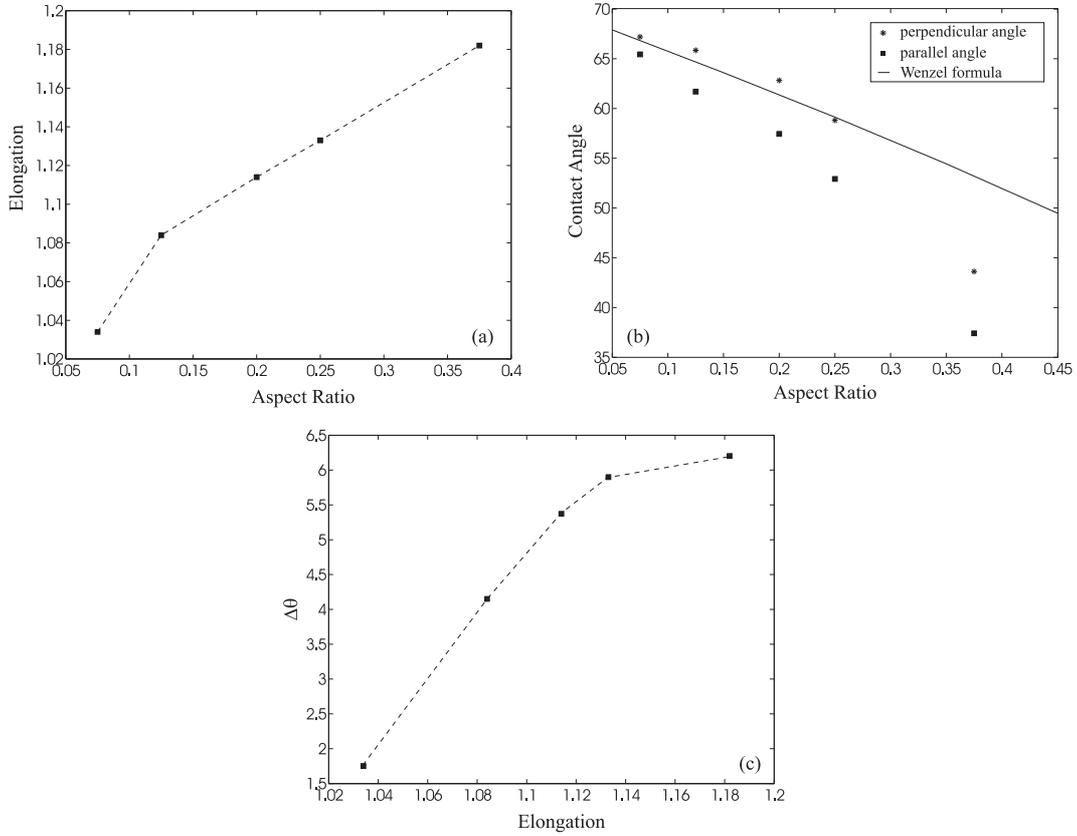}
    \caption{\footnotesize Lattice Boltzmann simulation results for drops on a corrugated surface. (a) Elongation of 
the droplet  and (b) $\theta_{\parallel}$ and $\theta_{\perp}$ as function of the aspect ratio of the surface. The 
elongation increases monotonically with increasing aspect ratio, but both $\theta_{\parallel}$ and $\theta_{\perp}$ 
decrease. The straight line corresponds to the Wenzel formula. (c) $\Delta\theta$ as a function of the elongation.
(The dotted lines in (a) and (c) are guides to the eye.)}
    \label{NumElongPerpParAspRatio}
    \end{figure}

The simulation results are presented in Table \ref{tab1} and Fig. \ref{NumElongPerpParAspRatio}. In interpreting the 
data it is important to note that there were two differences to the experiments forced upon us by computational 
requirements. Firstly, the dimensions of the simulated drop were comparable to the dimension of the grooves so that a drop lies 
on only 4 to 5 barriers. Secondly, the ridges were taken to be rectangular so that the maximum slope $\alpha = 
90^{\mathrm{o}}$ was independent of the aspect ratio. We varied the groove depth but kept the other parameters the 
same. In the simulations
    \begin{equation}
    r = 1 + \frac{2 \, h}{p} \, .
    \end{equation}

    \begin{table}
    \begin{center}
    \begin{tabular}{cccccc}
    \hline
    $w_1$  & $h$ & $\theta_{\perp}$ & $\theta_{\parallel}$ & $\theta_{\mathrm{Wenzel}}$ & $e$ \\
    \hline \hline
    \,\,\,\, 10 \,\,\,\,& \,\,\,\, 3 \,\,\,\, & \,\,\,\, 67.19 \,\,\,\, & \,\,\,\, 65.44 \,\,\,\, & \,\,\,\, 66.84 
\,\,\,\, & \,\,\,\, 1.034 \,\,\,\, \\
    10  &    5   &    65.85  &   61.70    &  64.69  &   1.084 \\
    10  &    8   &    62.83  &   57.45    &  61.39  &   1.114 \\
    10  &    10  &    58.82  &   52.92    &  59.13  &   1.133 \\
    10  &    15  &    43.61  &   37.41    &  53.23  &   1.182 \\
    \hline
    \end{tabular}
    \caption{\footnotesize Drop contact angles and elongation for different groove heights. The drop volume is $\sim 
7.47 \, \times \, 10^5$ and the groove width and pitch are
    kept constant at $30$ and $40$ respectively. $\theta_{\mathrm{Wenzel}}$ is the theoretical Wenzel angle. Here the 
drops only lie on 4 grooves.}
    \label{tab1}
    \end{center}
    \end{table}

Of the experimental features listed in the previous subsection, (iii) $\Delta\theta$ increases with increasing 
elongation $e$, and (iv) the drop elongation increases with increasing aspect ratio, are reproduced in the 
simulations, as shown in Fig. \ref{NumElongPerpParAspRatio}(a) and (c). These trends are observed for the same 
reasons as in the experiments.
         
\begin{figure}
    \centering
    \includegraphics[width=0.45\textwidth]{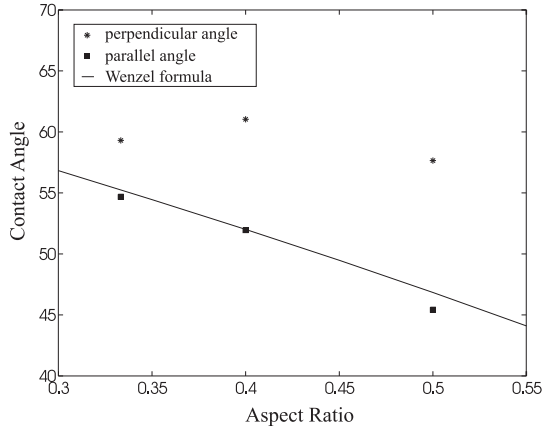}
    \caption{\footnotesize $\theta_{\parallel}$ and $\theta_{\perp}$ as a function of the aspect ratio of the 
surface. Here the size of the ridges is much smaller than
    the size of the drop. The solid line is the theoretical Wenzel angle.}
    \label{NumElongPerpParAspRatio2}
    \end{figure}

The experimental trends (i) and (ii) are, however, not reproduced by the simulations for the following reasons:

(i) $\theta_{\parallel} \simeq \theta_{\mathrm{Wenzel}}$: in the simulations (shown in Fig. \ref{NumElongPerpParAspRatio}(b)), 
the Wenzel angle is a poor approximation for the parallel contact angle, even though both simulations and experiments
are in the same aspect ratio regime. The reason for this discrepancy is that the size of the heterogeneities is of the same 
order as the size of the drop. Since the drop only lies on a small number of grooves, the Wenzel equation is not applicable. 
For this reason we performed a second set of simulations where the size of the ridges is an order of 
magnitude smaller than the drop, as shown in Table \ref{tab2} and Fig. \ref{NumElongPerpParAspRatio2}, and found 
the Wenzel equation is indeed a reasonable approximation to the parallel contact angle. In this case, 
we note that the size of the ridges is comparable to the width of the interface in the lattice Boltzmann simulations and
the roughness factor is much higher than that obtained in the experiments. In other words, as long as the typical length scale of
the corrugation is much smaller than the drop size, the Wenzel equation is a good estimate of the parallel contact angle.

    \begin{table}
    \begin{center}
    \begin{tabular}{cccccc}
    \hline
    $w_1$  & $p$ & $\theta_{\perp}$ & $\theta_{\parallel}$ & $\theta_{\mathrm{Wenzel}}$ & $e$ \\
    \hline \hline
    \,\,\,\, 4 \,\,\,\,& \,\,\,\, 8 \,\,\,\, & \,\,\,\, 57.64 \,\,\,\, & \,\,\,\, 45.42 \,\,\,\, & \,\,\,\, 46.84 
\,\,\,\, & \,\,\,\, 1.314 \,\,\,\, \\
    6  &    10   &    61.03  &   51.95    &  52.00  &   1.210 \\
    8  &    12   &    59.30  &   54.68    &  55.24  &   1.101 \\
    \hline
    \end{tabular}
    \caption{\footnotesize Drop contact angles and elongation for different groove widths. The drop volume is $\sim 
7.42 \, \times \, 10^5$ and the barrier width and height are
    kept constant at $4$ and $4$ respectively. $\theta_{\mathrm{Wenzel}}$ is the theoretical Wenzel angle.}
    \label{tab2}
    \end{center}
    \end{table}

(ii) $\theta_{\perp} > \theta_{\parallel}$ and $\theta_{\perp}$ on average increases with increasing aspect ratio: 
from Fig. \ref{NumElongPerpParAspRatio}(b) we indeed found that $\theta_{\perp} > \theta_{\parallel}$, however, from 
the data we have so far, increasing the aspect ratio is not followed by an increase in $\theta_{\perp}$; instead a 
decrease was found. We now argue why this may be the case. In the experiments, the roughness factor is close to 1 (at most 1.06, from equation \ref{eqRPE}) and 
hence $\theta_{\mathrm{\parallel}}$ is typically just below $\theta_e$. Furthermore, $\theta_A$ increases with the 
aspect ratio since $\alpha$ increases with the aspect ratio. Since $\theta_{\mathrm{\parallel}} < \theta_{\perp} < 
\theta_A$, $\theta_{\perp}$ increases with increasing aspect ratio. In the simulations, on the other hand, $r$ is 
considerably larger than 1 (between 1.15 and 2) and hence $\theta_{\mathrm{\parallel}}$ is much smaller than $\theta_e$. Since the lower 
limit of the allowed values of $\theta_{\perp}$ is decreased, it is plausible that we find $\theta_{\perp}$ 
decreasing with increasing aspect ratio in Fig. \ref{NumElongPerpParAspRatio}(b).


\section{Conclusions}

We have used experiments and simulations to investigate the behaviour of drops on surfaces patterned with sub-micron and
micron-scale parallel grooves. We find that the final drop shape is highly dependent on the path by which it is 
achieved. Drops which advance across the surface are elongated parallel to the grooves whereas drops that dewet the 
surface are elongated perpendicular to the grooves. We explain this behaviour in terms of the pinning of the contact 
line on the groove edges. We stress that it is not possible to measure a single contact angle for a drop on a 
patterned surface. The contact angle varies around the rim of the drop, with its position on the surface and with its 
dynamic history.

Contact angles and base dimensions were measured for drops spreading on corrugated surfaces. This was done for a 
large number of drops and substrate geometries. The data was very noisy, underlining the prevalence of hysteresis on
patterned surfaces. However, we were able to draw several conclusions from the results. We find that the parallel 
contact angle is close to the Wenzel angle. The Wenzel angle in return depends on the roughness factor of the surface.
Pinning effects render results for the perpendicular contact angles very noisy but they are, in general, greater than the 
parallel angles and tend to increase with increasing aspect ratio. The difference between parallel and perpendicular 
contact angles increases with the drop elongation whilst the drop elongation increases with the aspect ratio. This 
behaviour is explained, again by considering contact line pinning, and by noting that the drop spreads more quickly in 
the parallel direction on surfaces with higher aspect ratio.

We found that lattice Boltzmann simulations of drop motion were of great use in interpreting the experiments. The 
same relation between the direction of elongation of the drop and its direction of motion over the surface was 
immediately apparent in the simulations, and we were able to probe the pinning perpendicular to the grooves. 
Differences in detail between the simulations and experiments arose because the simulations are limited to drops 
lying on a small number of grooves. In particular, we find that the Wenzel angle no longer provides a good estimate 
of the parallel angle when the drop dimensions are comparable to the dimensions of the corrugation. An interesting 
future project would be to perform experiments on wider grooves where a quantitative match to the simulations may 
be possible. 


\section*{Acknowledgements}
(RV) Ko Hermans for practical help with photo-embossing, Marshall Ming and Paul van der Schoot for useful discussion. 
(HK) Matthew Blow and Ciro Semprebon for useful discussions. HK acknowledges support from a Clarendon Bursary and 
the INFLUS project.




\end{document}